\documentclass{article}
\usepackage[a4paper, margin=1in]{geometry} %
\usepackage{footnote}
\usepackage{rotating}
\usepackage{graphicx} 
\usepackage{color}
\usepackage{booktabs}
\usepackage{comment}
\usepackage{amsmath}
\usepackage{tabularx}
\usepackage{comment}
\usepackage{hyperref}
\usepackage{caption}
\usepackage{subfigure}
\usepackage{subcaption}
\usepackage{tablefootnote}
\makesavenoteenv{tabular}
\makesavenoteenv{table}
\usepackage{makecell}
\usepackage{authblk}
\usepackage{comment}
\usepackage{natbib}
\hypersetup{
	unicode=false,          
	pdftoolbar=true,        
	pdfmenubar=true,        
	pdffitwindow=false,     
	pdfstartview={FitH},    
	pdftitle={My title},    
	pdfauthor={Author},     
	pdfsubject={Subject},   
	pdfcreator={Creator},   
	pdfproducer={Producer}, 
	pdfkeywords={keywords}, 
	pdfnewwindow=true,      
	colorlinks=true,       
	linkcolor=blue,          
	citecolor=blue,        
	filecolor=magenta,      
	urlcolor=blue           
}

\title{Mathematical Modelling of Ethical AI Use in Higher Education: A Coordination Game Framework for Future-Facing Learning}

\author[1]{Ndidi Bianca Ogbo \thanks{Corresponding author: \texttt{b.ogbo@tees.ac.uk}}}
\author[1]{Zhao Song}
\author[1]{Shatha Ghareeb}
\author[1]{The Anh Han}
\affil[1]{School of Computing, Engineering and Digital Technologies, Teesside University, United Kingdom}

\begin{document}

\maketitle

\begin{abstract}
The rapid uptake of generative artificial intelligence (AI) in higher education is reshaping assessment practices and intensifying concerns around academic integrity, fairness, and learning quality. While institutional responses increasingly emphasise policy guidance and ethical principles, there remains limited formal understanding of how collective norms of responsible or opportunistic AI use emerge and stabilise within student cohorts. This paper reframes student AI use in assessment as a coordination problem shaped by peer expectations and assessment design rather than individual compliance alone. We develop a coordination-based evolutionary game-theoretic framework that captures learning value, effort, perceived fairness, and transparency, with institutional AI governance modelled implicitly through reflective assessment incentives.

We use analytical results and finite-population simulations to reveal threshold-driven behavioural transitions in student AI use: small, well-calibrated changes in reflective assessment incentives can trigger rapid shifts towards responsible, learning-oriented AI-use norms, whereas weak or misaligned incentives allow opportunistic practices to persist. These non-linear dynamics explain why policy statements alone often fail to change behaviour, while modest assessment redesigns can have disproportionate effects. By providing a mechanism-level account of how assessment structures shape collective AI-use practices, this work offers higher education institutions an analytically grounded tool for Future Facing Learning, supporting proportionate, pedagogy-led AI governance without reliance on surveillance or punitive enforcement.
\\\

\vspace{0.2em}
Evolutionary Game Theory
\end{abstract}

\section{Introduction}
\label{sec:introduction}

Artificial intelligence (AI) has become an increasingly influential component of higher education, shaping teaching, learning, assessment, and institutional decision-making. Over the past decade, AI-driven systems have supported personalised learning, automated feedback, learning analytics, and administrative processes \cite{Chen2020,Ouyang2022}. More recently, advances in generative AI have accelerated this transformation by enabling students to produce text, code, and other artefacts with unprecedented ease \cite{Selwyn2019,McDonald2025}. While these developments create significant pedagogical opportunities, they also raise pressing concerns regarding academic integrity, fairness, and the credibility of assessment \cite{Eaton2023,AlZahrani2024Ethical,powers_whats_2025,capraro2024impact}. In assessment contexts, generative AI tools can support exploration, drafting, and problem solving, yet they can also be used opportunistically to minimise effort or optimise short-term performance. Institutions therefore face a central challenge: how to design assessment environments that encourage responsible, learning-oriented AI use without undermining learning quality or perceived fairness \cite{Bearman2024,BerishaQehaja2025}. To date, institutional responses have focused largely on policy guidance, disclosure requirements, and detection-oriented approaches \cite{Chan2023,McDonald2025}. However, growing empirical evidence suggests that formal rules and enforcement mechanisms alone are insufficient to shape student behaviour in practice \cite{Perkins2024,Zhou2024}.

A key reason for this limitation is that student decisions about AI use are socially interdependent. Choices are shaped not only by institutional guidance, but also by peer behaviour, shared expectations, and perceptions of legitimacy within a cohort \cite{Ursavas2025,Ma2024}. In such environments, the costs and benefits of responsible or opportunistic AI use depend critically on what others are perceived to be doing. This interdependence gives rise to a coordination problem: once a particular pattern of AI use becomes widespread, it may be individually rational for students to conform, even when alternative behaviours are pedagogically preferable. Similar dynamics have been widely documented in studies of social norms and collective behaviour \cite{Merton1968,Sigmund2010}.

Recent scholarship highlights the role of assessment design in shaping AI-use practices, particularly through reflective and pedagogically aligned approaches that require students to articulate and critically evaluate how AI tools contributed to their work, and through task designs that emphasise higher-order thinking over surface-level outputs \cite{kizilcec2024perceptions,Khlaif2025Redesign,Urquhart2026Assessment}. Such approaches are increasingly advocated in policy and practice, yet there remains limited formal understanding of how reflective assessment mechanisms interact with peer dynamics to shape collective AI-use norms over time. In particular, it remains unclear under what conditions responsible AI-use norms stabilise, and when opportunistic or superficially compliant behaviours persist.

To address this gap, we introduce a coordination-based modelling framework for analysing how patterns of AI use emerge within higher education assessment environments. Building on evolutionary game theory \cite{Nowak2004,Traulsen2006}, we model student--student interactions to capture norm formation driven by peer comparison, an approach widely used to study coordination and technology-adoption dynamics \cite{ZhuWeyant2003,OgboElragigHan2022,OgboHan2019}. Institutional influence enters implicitly through assessment design, which shapes the payoff structure by determining which forms of AI use are rewarded or discouraged.

Our analysis proceeds in two stages. First, we introduce a baseline coordination model capturing the tension between responsible, learning-oriented AI use and opportunistic, performance-driven use. We then extend this model by incorporating a reflective assessment mechanism that differentiates between meaningful and superficial engagement with AI. Using evolutionary dynamics in finite populations, we examine how these design choices influence the emergence, stability, and transition between collective AI-use norms within a student cohort.

By treating assessment design as a coordination mechanism rather than a compliance tool, this work contributes to Future Facing Learning by offering scalable, pedagogy-led approaches to AI governance in higher education. More broadly, it provides an analytically grounded perspective on how modest changes in assessment design can generate non-linear shifts in collective behaviour, offering practical insights for institutions seeking to support responsible, trustworthy, and pedagogically coherent uses of generative AI.

This paper makes three contributions. First, it introduces a coordination-based evolutionary game-theoretic framework for understanding how student AI-use practices are shaped by peer expectations. Second, it formalises reflective assessment design as an institutional mechanism that reshapes assessment-related incentives. Third, it provides analytical and computational insights into how responsible, superficially compliant, and opportunistic AI-use norms emerge and stabilise under different assessment designs, with implications for institutional AI governance and future-facing assessment practice.

\section{Background and Related Work}
\label{sec:background}

The rapid diffusion of artificial intelligence, and particularly generative AI tools, has intensified debate within higher education around pedagogy, assessment, and academic integrity. Recent research spans empirical studies of student and staff perceptions, institutional policy analyses, and conceptual frameworks for responsible AI use. While this literature provides important descriptive and normative insights, it offers limited analytical understanding of how collective norms of AI use emerge, stabilise, or shift within student cohorts, particularly in response to assessment design.

Early work on artificial intelligence in education focused on intelligent tutoring systems, adaptive learning, and learning analytics \cite{Chen2020}. More recent studies document the widespread uptake of generative AI tools in higher education and highlight both pedagogical opportunities and integrity-related concerns \cite{Selwyn2019,McDonald2025}. Systematic reviews consistently emphasise that educational outcomes depend less on the technical capabilities of AI systems than on how they are embedded within teaching and assessment practices \cite{Ouyang2022}. This shift has redirected attention away from technology itself toward institutional design choices that shape how AI is used in practice. At the institutional level, analyses of policy documents and guidelines reveal a move away from prohibition toward principles of responsible use, transparency, and AI literacy \cite{Chan2023,McDonald2025}. However, these responses remain largely normative and aspirational, offering limited insight into how students respond strategically to assessment incentives or how shared expectations around acceptable AI use form and persist within cohorts.

A growing body of empirical work helps to explain this gap by demonstrating that student AI-use decisions are socially interdependent. Quantitative studies drawing on technology acceptance models identify perceived usefulness, ease of use, and social influence as key drivers of AI adoption \cite{Chatterjee2020,Tang2024,Hussain2025}. Complementary qualitative and cross-cultural studies show that perceptions of peer behaviour and legitimacy often outweigh formal rules in shaping how students engage with AI tools \cite{Ma2024,Zhou2024,Ursavas2025}. Together, these findings suggest that AI use in assessment is not simply an individual compliance decision, but a coordination problem in which the relative attractiveness of responsible or opportunistic behaviour depends on what others are perceived to be doing.

Assessment design plays a central role in structuring this coordination problem. Long-standing work in higher education demonstrates that assessment strongly shapes how students allocate effort, interpret learning priorities, and define what constitutes acceptable performance \cite{BoudFalchikov2007}. In the context of generative AI, recent scholarship argues that approaches centred on detection or rule enforcement are insufficient, advocating instead for assessment designs that foreground evaluative judgement, reflection, and ethical engagement with AI tools \cite{Eaton2023,Bearman2024}. Reflective assessment tasks are increasingly proposed as a means of encouraging responsible AI use by making learning processes visible rather than merely policing outputs. However, existing studies largely treat reflection as an individual-level intervention and do not formally analyse how reflective assessment incentives interact with peer dynamics to reshape collective behaviour. In particular, prior research on student reflection, assessment design, and academic integrity does not explain why reflective incentives may generate rapid norm shifts in some contexts, gradual shifts in others, or fail altogether when reflective effort is poorly calibrated, nor does it analytically distinguish between meaningful reflection and superficial or symbolic compliance.

Game-theoretic and evolutionary approaches offer a natural framework for addressing these limitations \cite{axelrod84,novak2006,Sigmund2010,ALALAWI2026129627,ogbo2022shake,NdidiZhaoTheAnh2025}. Coordination games and evolutionary dynamics have been widely used to study norm formation, technology adoption, and the role of institutional mechanisms in shaping collective outcomes. Prior work demonstrates how incentive structures, social learning, and institutional mechanisms such as prior commitments can transform strategic environments and resolve inefficient equilibria \cite{OgboHan2019,OgboElragigHan2022,NBOgbo,han2022Interface}. While related modelling approaches are beginning to appear in the context of AI and education \cite{Yuan2025,You2023}, they typically operate at an abstract stakeholder level and do not examine assessment-level decision-making, peer-driven norm formation, or qualitatively distinct forms of AI use within student cohorts.

Against this backdrop, there remains a clear need for analytical frameworks that connect assessment design, peer expectations, and student behaviour within a unified model. By building on evolutionary coordination theory and grounding it in contemporary debates on generative AI and assessment, this paper develops a formal framework for analysing how responsible, superficially compliant, and opportunistic patterns of AI use emerge, compete, and stabilise in higher education assessment contexts.

\section{Model and Methods}
\label{sec:model}

We develop a coordination-based modelling framework to analyse how patterns of generative AI use emerge within higher education assessment environments. The framework captures how individual student decisions interact with peer expectations to generate shared norms of AI use within a cohort, with institutional influence entering implicitly through assessment design. Existing research indicates that student engagement with generative AI is shaped not only by formal rules, but also by perceptions of fairness, legitimacy, peer behaviour, and the signals conveyed by assessment tasks regarding acceptable forms of AI use \cite{Selwyn2019,Bearman2024,Perkins2024}. Accordingly, we adopt a norm-based perspective in which assessment design structures the strategic environment in which student behaviour evolves, consistent with long-standing evidence that assessment strongly shapes learner behaviour and study practices \cite{BoudFalchikov2007}, as well as recent arguments for framing responsible AI engagement in terms of learning and ethics rather than narrow compliance \cite{Eaton2023,Bearman2024}. 

In particular, we model these interactions using evolutionary game-theoretic methods in finite populations \cite{traulsen2006evolution,novak2006,Sigmund2010}, allowing us to analyse how peer interaction, assessment incentives, and social learning jointly shape the emergence and stability of AI-use norms over time.

\subsection{Decision Context and Modelling Approach}

We consider a cohort of students enrolled in a course where generative AI tools are available for learning and assessment-related activities. Each assessment task is modelled as a \emph{one-shot decision context}, in which students choose how to engage with AI tools. Although decisions are one-shot at the individual level, students repeatedly encounter similar assessment environments and observe peer practices and outcomes over time. These repeated exposures contribute to the formation of shared expectations and norms regarding acceptable AI use, consistent with sociological accounts of norm emergence and social regulation \cite{Merton1968}.

To capture this setting in a tractable way, we model interactions at the student--student level using a two-player symmetric game in a well-mixed population. Each player represents a typical student randomly matched with another student from the same cohort. The institution does not appear as an explicit strategic player; instead, it shapes behaviour indirectly through assessment design, which determines the payoff structure associated with different forms of AI use.

\subsection{Baseline Model: Responsible and Opportunistic AI Use}

We begin with a minimal baseline model capturing the core behavioural tension in AI-supported assessment: the trade-off between responsible, learning-oriented AI use and opportunistic, performance-driven AI use \cite{Perkins2024}. Students are assumed to balance perceived learning value against time pressure, workload, and performance optimisation when deciding how to use AI tools. We therefore consider two strategies:
\begin{itemize}
    \item \textbf{R (Responsible AI use):} AI is used in ways aligned with assessment rules and learning objectives.
    \item \textbf{O (Opportunistic AI use):} AI is used primarily to reduce effort or optimise short-term performance.
\end{itemize}

\begin{table}[t!]
\centering
\caption{List of parameters in the AI-use coordination model.}
\label{tbl:AIModel}
\begin{tabular}{|p{9cm}|c|}
\hline
\textbf{Parameter description} & \textbf{Notation} \\
\hline \hline
Additional learning benefit from responsible AI use & $L$ \\
Effort cost associated with responsible AI use & $C$ \\
Short-term performance advantage of opportunistic AI use & $S$ \\
Norm or legitimacy cost from misalignment with peers & $\delta$ \\
Misconduct cost associated with strategy $M$ & $\tau$ \\
Assessment reward for meaningful reflective engagement & $r$ \\
Effort cost of meaningful reflective engagement & $\kappa$ \\
Fraction of reward and effort for superficial reflection & $\sigma$ \\
Population size & $N$ \\
Intensity of social learning & $\beta$ \\
Exploration (mutation) probability & $\mu$ \\
\hline
\end{tabular}
\end{table}

\subsubsection{Baseline Payoff Structure}

Let $\pi_{i,j}$ denote the payoff to a student using strategy $i$ when interacting with a peer using strategy $j$. Since evolutionary dynamics under pairwise comparison depend only on payoff differences, any common baseline term can be removed without loss of generality \cite{Nowak2004,Traulsen2006}. The interaction between responsible and opportunistic AI use is therefore represented by
\begin{equation}
\label{eq:AI_abcd}
\bordermatrix{~ & R & O \cr
R & L - C & L - C - \delta \cr
O & S - \delta & S \cr
}.
\end{equation}
Responsible AI use yields additional learning value $L$ at an effort cost $C$, while opportunistic AI use provides a short-term performance advantage $S$. When students adopt different behaviours, a legitimacy cost $\delta$ is incurred due to misalignment with cohort norms, reflecting perceived unfairness or free-riding concerns documented in assessment practice \cite{Bearman2024,Ajjawi2020}. The payoff structure satisfies $a>b$ and $d>c$, defining a coordination game in which alignment with peer behaviour is individually advantageous.

\subsection{Institutional Mechanism: Reflective Assessment Design}

We extend the baseline model by introducing reflective engagement with AI use, an institutional mechanism increasingly advocated in higher education. Rather than relying on detection or enforcement, reflective assessment requires students to articulate, justify, and critically evaluate how generative AI tools contributed to their work \cite{Bearman2024,BoudFalchikov2007,MolloyBoudHenderson2020}.

Reflective requirements differentiate patterns of student behaviour beyond the simple distinction between responsible and opportunistic AI use. We therefore consider four strategies, summarised in Table~\ref{tab:ai_strategies}. Strategy $O$ represents opportunistic but ambiguously tolerated AI use, whereas $M$ represents clear misuse that violates assessment rules and incurs additional penalties.

\begin{table}[t]
\centering
\caption{Strategy set describing patterns of AI use in assessment.}
\label{tab:ai_strategies}
\begin{tabular}{p{1.6cm}||p{3.2cm}|p{3.4cm}|p{3.2cm}}
\hline
\textbf{Strategy} & \textbf{Responsible AI use?} & \textbf{Reflective engagement} & \textbf{Rule compliance} \\
\hline \hline
RR & Yes & Meaningful reflection & Yes \\
RS & Yes & Superficial reflection & Yes \\
O  & No  & No reflection & Ambiguous \\
M  & No  & No reflection & No \\
\hline
\end{tabular}
\end{table}

The resulting interaction is captured by the payoff matrix
\begin{equation}
\label{eq:AI_4x4}
\bordermatrix{~ & RR & RS & O & M \cr
RR & a + r - \kappa & a + r - \kappa & b + r - \kappa & b - \kappa \cr
RS & a + \sigma r - \sigma \kappa & a + \sigma r - \sigma \kappa & b + \sigma r - \sigma \kappa & b - \sigma \kappa \cr
O  & c - \delta & c - \delta & d & d \cr
M  & c - \delta - \tau & c - \delta - \tau & d - \tau & d - \tau \cr
}.
\end{equation}

Meaningful reflection yields a reward $r$ at effort cost $\kappa$, while superficial reflection yields a fraction $\sigma r$ at reduced cost $\sigma\kappa$. Clear misuse incurs an additional misconduct cost $\tau$. This structure preserves the coordination logic of the baseline model while distinguishing meaningful engagement from superficial compliance.

\subsection{Evolutionary Dynamics}

We analyse behavioural dynamics using a finite-population evolutionary game-theoretic framework in which payoffs represent relative success within the assessment environment. Behavioural change occurs through social learning, whereby students compare outcomes and adopt strategies that appear more successful.

Consider a well-mixed population of size $N$. If $k$ students use strategy $i$ and $N-k$ use strategy $j$, the average payoffs are
\begin{equation}
\label{eq:avg_payoff}
\begin{split}
\Pi_i(k) &= \frac{(k-1)\pi_{ii} + (N-k)\pi_{ij}}{N-1}, \\
\Pi_j(k) &= \frac{k\,\pi_{ji} + (N-k-1)\pi_{jj}}{N-1}.
\end{split}
\end{equation}

Strategy updating follows the Fermi imitation rule
\begin{equation}
\label{eq:fermi}
p_{i \rightarrow j}
=
\left(1 + e^{-\beta\left(\Pi_j(k) - \Pi_i(k)\right)}\right)^{-1},
\end{equation}
where $\beta$ denotes the intensity of social learning \cite{Traulsen2006,Sigmund2010}. A small mutation probability $\mu$ allows occasional exploration of alternative strategies.

For sufficiently small $\mu$, the population spends most of its time in homogeneous states. The long-run dynamics are approximated by a Markov chain over these states, with fixation probability
\begin{equation}
\label{eq:fixation}
\rho_{j,i}
=
\left(
1 + \sum_{m=1}^{N-1}
\prod_{\ell=1}^{m}
\frac{T^{-}(\ell)}{T^{+}(\ell)}
\right)^{-1},
\end{equation}
where
\begin{equation}
\label{eq:transition}
T^{\pm}(m)
=
\frac{m}{N}\frac{N-m}{N}
\left(
1 + e^{\mp \beta\left(\Pi_i(m) - \Pi_j(m)\right)}
\right)^{-1}.
\end{equation}

The stationary distribution of this Markov chain characterises the proportion of time the population spends in each homogeneous AI-use regime \cite{Imhof2005,Traulsen2006,Nowak2004}. This framework allows us to identify evolutionarily stable patterns of AI use and to examine how assessment design parameters induce non-linear transitions between responsible, superficially compliant, and opportunistic norms within a student cohort.

\section{Results}
\label{sec:result}

We present results from numerical simulations, based on the evolutionary dynamics described in Section~\ref{sec:model}, examining how assessment design parameters shape the emergence and stability of collective AI-use norms. We focus on the role of reflective engagement incentives in shaping long-run evolutionary stability across competing AI-use strategies, highlighting conditions under which responsible, transparent practices emerge as dominant social norms.

\subsection{Emergence of Responsible AI Use Under Reflective Assessment Design}
\label{subsec:reflection}

\begin{figure*}[htbp]
\centering
\includegraphics[width=0.7\linewidth]{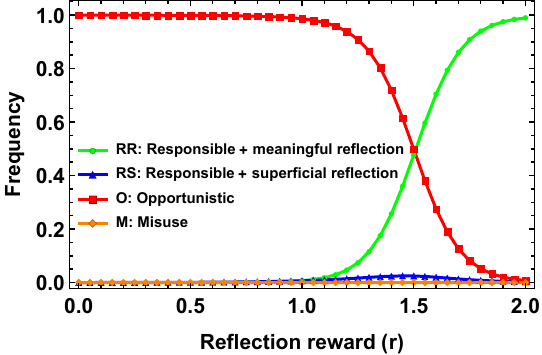}
\caption{Stationary frequencies of AI-use strategies as a function of the reflection reward $r$.
The figure shows the long-run frequencies of responsible AI use with meaningful reflection (RR), responsible AI use with superficial reflection (RS), opportunistic AI use (O), and misuse (M).
Stationary frequencies are computed from a small-mutation Markov chain over homogeneous states using fixation probabilities under Fermi updating.
Parameters: $a = 1$, $b = 0$, $c = 1$, $d = 2$; reflection effort cost $\kappa = 1$; superficial reflection factor $\sigma = 0.4$; legitimacy cost $\delta = 1$; misuse penalty $\tau = 1$; population size $N = 100$; $\beta = 0.1$.}
    \label{fig:transition_r}
\end{figure*}

Figure~\ref{fig:transition_r} shows the stationary frequencies of four AI-use strategies as the reflection reward $r$ increases.
For low values of $r$, opportunistic AI use dominates the population, with responsible practices occurring only at negligible levels.
This regime reflects assessment environments in which reflective engagement is weakly incentivised and AI use remains largely instrumental and unregulated by shared norms.

As the reflection reward increases, a sharp transition occurs around a critical threshold of $r \approx 1.5$.
Beyond this point, responsible AI use coupled with meaningful reflection rapidly becomes the dominant strategy, while opportunistic AI use correspondingly collapses.
The abrupt nature of this transition indicates a coordination-driven regime shift, in which behavioural success depends on alignment with prevailing cohort practices rather than individual optimisation alone.

Across the entire range of $r$, responsible AI use with superficial reflection remains at consistently low frequency.
This suggests that symbolic or minimally embedded reflective activities are insufficient to displace opportunistic behaviour or establish stable responsible norms.
Misuse also remains negligible throughout, indicating that the primary governance challenge in higher education lies not in overt misconduct, but in the widespread normalisation of opportunistic yet superficially acceptable AI use.

Overall, these results demonstrate that reflective assessment design can function as a structural coordination mechanism.
By embedding meaningful reflection as a valued component of assessment, institutions can shift collective behaviour away from opportunistic AI use and towards responsible, transparent, and pedagogically aligned engagement.

\subsection{Peer sensitivity determines how rapidly responsible AI-use norms emerge under reflective assessment}
\label{subsec:peersensitivity}

\begin{figure*}[htbp]
\centering
\includegraphics[width=0.7\linewidth]{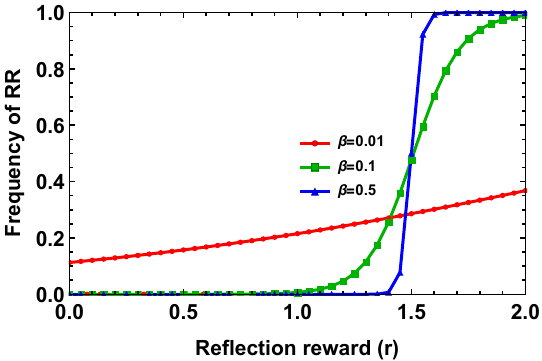}
\caption{
Frequency of responsible AI use (RR) as a function of $r$ (reflection reward) under different levels of peer sensitivity.
The stationary frequency of RR is shown for $\beta = 0.01$, $0.1$, and $0.5$. Payoff parameters: $a = 1$, $b = 0$, $c = 1$, $d = 2$; $\delta = 1$; $\kappa = 1$; $\sigma = 0.4$; $\tau = 1$.
Other parameters: population size $N = 100$.
}
\label{fig:sensitivity}
\end{figure*}

Figure~\ref{fig:sensitivity} illustrates how the effectiveness of reflective assessment in promoting responsible AI use depends critically on peer sensitivity within student cohorts. While increases in reflection reward encourage responsible behaviour across all scenarios, the speed and stability of norm formation vary substantially depending on how strongly students respond to peer practices and shared expectations.

When peer sensitivity is weak ($\beta = 0.01$), responsible AI use increases only gradually with higher reflection rewards, remaining marginal even at high incentive levels. This regime corresponds to learning environments in which AI use is largely individualised and weakly shaped by social norms. At moderate peer sensitivity ($\beta = 0.1$), responsible AI use emerges more reliably as incentives increase, reflecting contexts where students observe and compare AI practices but norms remain fluid.

Under strong peer sensitivity ($\beta = 0.5$), the system exhibits a sharp transition: once a critical reflection threshold is reached, responsible AI use rapidly becomes the dominant norm. This tipping-point behaviour reflects norm cascades commonly observed in educational settings, where visible practices, shared standards, and collective discussion accelerate behavioural alignment.

\subsection{Assessment design constraints determine when responsible AI-use norms are feasible}
\label{subsec:designspace}

\begin{figure*}[htbp]
\centering
\includegraphics[width=0.7\linewidth]{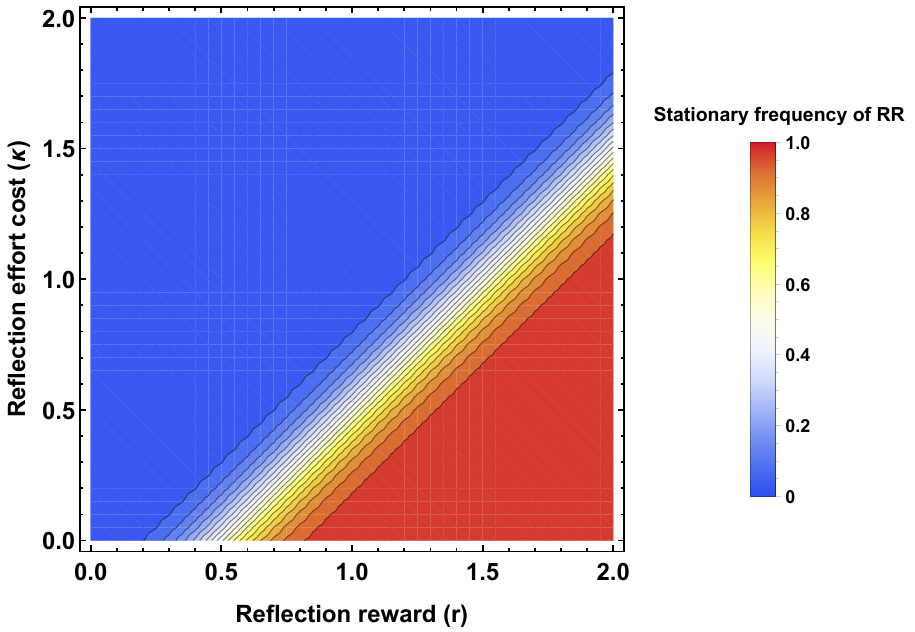}
\caption{\textbf{Design space of reflective AI assessment.}
Stationary frequency of responsible AI use with meaningful reflection (RR) as a function of reflection reward $r$ and reflection effort cost $\kappa$. Warmer colours indicate regimes in which RR dominates in the long run, while cooler colours indicate dominance of opportunistic AI use. Parameters: $a=1$, $b=0$, $c=1$, $d=2$, $\delta=1$, $\sigma=0.4$, $\tau=1$, population size $N=100$, and peer sensitivity $\beta=0.1$.}
\label{fig:designspace}
\end{figure*}

Figure~\ref{fig:designspace} illustrates how the long-run emergence of responsible AI-use norms depends jointly on the calibration of reflective rewards and the effort required to engage in reflection. Rather than varying incentives in isolation, the figure maps the feasible assessment design space within which reflective assessment can support responsible and pedagogically aligned AI use.

When reflection effort is high relative to the reward assigned to it (upper-left region), responsible AI use remains rare, even when reflective incentives are present. In this regime, the cognitive and time demands associated with meaningful reflection outweigh its perceived assessment value, making opportunistic AI use the evolutionarily stable outcome. This finding reflects longstanding observations in higher education that students respond strategically to assessment demands, particularly when tasks are perceived as burdensome or misaligned with credit allocation \cite{bearman2022assessment}.

As reflection rewards increase or reflection effort decreases (moving towards the lower-right region), responsible AI use with meaningful reflection becomes increasingly prevalent and ultimately dominates the population. This regime corresponds to assessment designs in which reflective engagement is both pedagogically meaningful and credibly valued. Under these conditions, responsible AI use stabilises not as an exceptional behaviour, but as a shared social norm supported by assessment structure.

Between these regimes lies a narrow transition region characterised by a sharp shift in outcomes. Small changes in reward or effort near this boundary produce large changes in the stationary frequency of responsible AI use, indicating a tipping-point dynamic. This suggests that assessment-led AI governance is highly sensitive to calibration: modest adjustments to reflective weighting or task complexity can determine whether responsible AI use remains marginal or becomes the dominant collective practice.

Overall, Figure~\ref{fig:designspace} demonstrates that reflective assessment is effective only within a constrained design space. Responsible AI-use norms emerge not simply from increasing incentives, but from aligning reflective effort with assessment value. This result provides a formal, quantitative foundation for recent calls within higher education to address generative AI through assessment redesign and proportional reflective practices, rather than through increased surveillance or punitive controls \cite{dawson2023assessmentAI}.

\section{Discussion and Conclusion}
\label{sec:discussion}

This paper examined how assessment design shapes the emergence of responsible AI-use norms in higher education. By modelling student engagement with generative AI as a coordination problem influenced by institutional incentives and peer dynamics, the analysis explains why responsible practices may fail to stabilise despite clear ethical guidance or institutional policy.

Across Figures~1–3, a consistent pattern emerges: responsible AI use does not increase smoothly with incentives, but instead arises through threshold-driven transitions. Figure~1 shows that embedding reflection as a meaningful and rewarded component of assessment can displace opportunistic AI use, but only once a critical reward level is reached. Below this threshold, opportunistic behaviour remains dominant, reflecting students’ strategic responses to assessment structures rather than a lack of ethical intent. This aligns with long-standing evidence that assessment design strongly shapes learning behaviour and academic norms \cite{BoudFalchikov2007,bearman2022assessment}.

Figure~2 demonstrates that peer sensitivity plays a crucial role in determining how rapidly responsible AI-use norms emerge. When students are highly responsive to peer practices, relatively modest changes in assessment incentives can trigger rapid norm cascades. In contrast, weak peer sensitivity leads to gradual and incomplete transitions. These dynamics are consistent with empirical findings showing that perceptions of peer behaviour and shared expectations strongly influence how students interpret the legitimacy of AI use in assessment contexts \cite{kizilcec2024perceptions}.

Figure~3 identifies structural constraints on reflective assessment design. Responsible AI use emerges most robustly when the reward for reflection is proportionate to the effort required to engage in it. When reflective tasks impose high cognitive or time demands without sufficient assessment value, opportunistic AI use persists even under relatively generous incentives. This provides a formal explanation for why well-intentioned but overly burdensome assessment interventions may be ineffective, complementing systematic evidence that generative AI primarily exposes weaknesses in traditional assessment design rather than causing integrity problems per se \cite{bittle2025genAIreview}.

Taken together, these results suggest that effective AI governance in higher education is fundamentally an assessment design challenge rather than a compliance problem. Survey-based evidence indicates that both educators and students support redesign approaches that emphasise transparency, reflection, and authentic engagement when assessment tasks are credibly aligned with learning goals \cite{kizilcec2024perceptions}. Our model provides a mechanism-level explanation for these perceptions, showing how proportional incentives and social learning can stabilise responsible AI-use norms without reliance on detection or punitive controls.

More broadly, this work connects debates on AI in higher education with insights from evolutionary coordination theory. As in other cooperative settings, poorly calibrated incentives can generate hidden costs and fail to improve collective outcomes \cite{Han2026CooperationWelfare}. Framing assessment design as a coordination mechanism rather than a policing tool highlights how modest pedagogical changes can produce non-linear shifts in collective behaviour.

Several extensions would strengthen the framework’s explanatory and policy relevance. 
First, extending the model to multi-player and networked interactions would better capture collaborative learning, peer dialogue, and group-based reflection in higher education, enabling analysis of how group size and structure shape responsible AI-use norms \cite{NBOgbo,Bearman2024,ogbo2022shake}. 
Second, alternative incentive mechanisms such as reward-based, penalty-based, and hybrid designs could be compared in terms of institutional cost and social welfare, building on recent evolutionary analyses of incentive efficiency and collective outcomes \cite{Han2026CooperationWelfare}. 
Finally, empirical validation through surveys, behavioural experiments, or interactive assessment settings would help ground the model’s predictions in observed student behaviour and disciplinary contexts \cite{kizilcec2024perceptions,bittle2025genAIreview}.

\clearpage
\bibliographystyle{unsrt}
\bibliography{references}

@inproceedings{ogbo2022shake,
  title={Shake on it: The role of commitments and the evolution of coordination in networks of technology firms},
  author={Ogbo, Ndidi Bianca and Cimpeanu, Theodor and Di Stefano, Alessandro and Han, The Anh},
  booktitle={Artificial Life Conference Proceedings 34},
  volume={2022},
  number={1},
  pages={41},
  year={2022},
  organization={MIT Press}
}

@article{ALALAWI2026129627,
title = {Trust AI regulation? Discerning users are vital to build trust and effective AI regulation},
journal = {Applied Mathematics and Computation},
volume = {508},
pages = {129627},
year = {2026},
issn = {0096-3003},
doi = {https://doi.org/10.1016/j.amc.2025.129627},
url = {https://www.sciencedirect.com/science/article/pii/S0096300325003534},
author = {Zainab Alalawi and Paolo Bova and Theodor Cimpeanu and Alessandro {Di Stefano} and Manh {Hong Duong} and Elias Fernández Domingos and The Anh Han and Marcus Krellner and Ndidi Bianca Ogbo and Simon T. Powers and Filippo Zimmaro}
}

@article{capraro2024impact,
  title={The impact of generative artificial intelligence on socioeconomic inequalities and policy making},
  author={Capraro, Valerio and Lentsch, Austin and Acemoglu, Daron and Akgun, Selin and Akhmedova, Aisel and Bilancini, Ennio and Bonnefon, Jean-Fran{\c{c}}ois and Bra{\~n}as-Garza, Pablo and Butera, Luigi and Douglas, Karen M and others},
  journal={PNAS nexus},
  volume={3},
  number={6},
  pages={pgae191},
  year={2024},
  publisher={Oxford University Press US}
}

@article{powers_whats_2025,
	title = {What’s {It} {Like} to {Trust} an {LLM}: {The} {Devolution} of {Trust} {Psychology}?},
	volume = {44},
	issn = {1937-416X},
	shorttitle = {What’s {It} {Like} to {Trust} an {LLM}},
	url = {https://ieeexplore.ieee.org/document/11163569},
	doi = {10.1109/MTS.2025.3583233},
	number = {3},
	urldate = {2025-09-23},
	journal = {IEEE Technology and Society Magazine},
	author = {Powers, Simon T. and Urquhart, Neil and Barnes, Chloe M. and Cimpeanu, Theodor and Ekárt, Anikó and Han, The Anh and Pitt, Jeremy and Guckert, Michael},
	month = sep,
	year = {2025},
	pages = {30--37},
}

@article{han2022Interface,
  title={Institutional Incentives for the Evolution of Committed Cooperation: Ensuring Participation is as Important as Enhancing Compliance},
  author={Han, The Anh},
  journal={Journal of the Royal Society Interface},
  volume = {19},
number = {188},
pages = {20220036},
year = {2022},
doi = {10.1098/rsif.2022.0036}
}

@article{NBOgbo,
author = {Ndidi Bianca Ogbo and Aiman Elragig and The Anh Han},
title ={Evolution of coordination in pairwise and multi-player interactions via prior commitments},
journal = {Adaptive Behavior},
volume = {30},
number = {3},
pages = {257-277},
year = {2022},
doi = {10.1177/1059712321993166}
}

@article{ZhuWeyant2003,
  title={Strategic decisions of new technology adoption under asymmetric information: a game-theoretic model},
  author={Zhu, Kevin and Weyant, John P},
  journal={Decision sciences},
  volume={34},
  number={4},
  pages={643--675},
  year={2003},
  publisher={Wiley Online Library}
}

@article{traulsen2006evolution,
	Author = {Traulsen, Arne and Nowak, Martin A},
	Journal = {Proceedings of the National Academy of Sciences},
	Number = {29},
	Pages = {10952--10955},
	Publisher = {National Acad Sciences},
	Title = {Evolution of cooperation by multilevel selection},
	Volume = {103},
	Year = {2006}}

@article{Imhof2005,
	Author = {L. A. Imhof and D. Fudenberg and Martin A. Nowak},
	Journal = {Proc. Natl. Acad. Sci. U.S.A.},
	Pages = {10797--10800},
	Title = {Evolutionary cycles of cooperation and defection},
	Volume = {102},
	Year = {2005}}

@article{traulsen2006,
  title = {Stochastic dynamics of invasion and fixation},
  author = {Traulsen, Arne and Nowak, Martin A. and Pacheco, Jorge M.},
  journal = {Phys. Rev. E},
  volume = {74},
  issue = {1},
  pages = {011909},
  numpages = {5},
  year = {2006},
  month = {Jul},
  publisher = {American Physical Society},
  url ={https://link.aps.org/doi/10.1103/PhysRevE.74.011909}
}

@book{Sigmund2010,
	Author = {Karl Sigmund},
	Publisher = {Princeton University Press},
	Title = {The Calculus of Selfishness},
	Year = {2010}}

@book{axelrod84,
	Author = {Robert Axelrod},
	Publisher = {Basic Books, ISBN 0-465-02122-2},
	Title = {The Evolution of Cooperation},
	Year = {1984}}

@article{Nowak2004,
	Author = {M. A. Nowak and A. Sasaki and C. Taylor and D. Fudenberg},
	Journal = {Nature},
	Pages = {646--650},
	Title = {Emergence of cooperation and evolutionary stability in finite populations},
	Volume = {428},
	Year = {2004}}

@article{novak2006,
	Author = {Martin A. Nowak},
	Journal = {Science},
	Number = {5805},
	Pages = {1560},
	Title = {Five Rules for the Evolution of Cooperation},
	Volume = {314},
	Year = {2006}}

@inproceedings{OgboHan2019,
  title={Emergence of Coordination with Asymmetric Benefits via Prior Commitment},
  author={Bianca, Ogbo Ndidi and Han, The Anh},
  booktitle={Artificial Life Conference Proceedings},
  pages={163--170},
  year={2019},
  organization={MIT Press}
}

@inbook{OgboElragigHan2022,
  author       = {Ogbo, Ndidi Bianca and Han, The Anh},
  title        = {Coordination Dynamics in Technology Adoption: Lessons From an Evolutionary Game Theoretical Analysis},
  booktitle    = {Multisector Insights in Healthcare, Social Sciences, Society, and Technology},
  series       = {Multisector Insights in Healthcare, Social Sciences, Society, and Technology},
  publisher    = {IGI Global},
  year         = {2024},
  pages        = {295--326},
  doi          = {10.4018/979-8-3693-3226-9.ch016},
  isbn         = {9798369332269},
}

@book{Selwyn2019,
title = "Should robots replace teachers?: AI and the Future of Education",
author = "Neil Selwyn",
year = "2019",
month = nov,
language = "English",
isbn = "9781509528967",
publisher = "Polity Press",
address = "United Kingdom",
edition = "1st",
}

@article{Bearman2024,
author = {Margaret Bearman and Joanna Tai and Phillip Dawson and David Boud and Rola Ajjawi},
title = {Developing evaluative judgement for a time of generative artificial intelligence},
journal = {Assessment \& Evaluation in Higher Education},
volume = {49},
number = {6},
pages = {893--905},
year = {2024},
publisher = {Routledge},
URL = {https://doi.org/10.1080/02602938.2024.2335321
    },
eprint = {https://doi.org/10.1080/02602938.2024.2335321
        }
}

@misc{Perkins2024,
	author = 	 {Mike Perkins and Leon Furze and Jasper Roe and Jason Macvaugh},
	year = 	 {2024},
	month = 	 {-04-30},
	title = 	 {The Artificial Intelligence Assessment Scale ({AIAS}): A Framework for Ethical Integration of Generative {AI} in Educational Assessment},
	journal = 	 {Journal of University Teaching and Learning Practice},
	volume = 	 {21},
	number = 	 {06},
	isbn = 	 {1449-9789},
	doi={10.53761/q3azde36}
}

@misc{Eaton2023,
	author = 	 {Sarah Elaine Eaton},
	year = 	 {2023},
	month = 	 {-10-12},
	title = 	 {Postplagiarism: transdisciplinary ethics and integrity in the age of artificial intelligence and neurotechnology},
	journal = 	 {International Journal for Educational Integrity},
	volume = 	 {19},
	number = 	 {1},
	doi={10.1007/s40979-023-00144-1}
}

@book{BoudFalchikov2007,
	author={David Boud and Nancy Falchikov},
	year={2007},
	title={Rethinking assessment in higher education : learning for the longer term},
	publisher={Routledge},
	address={Abingdon, Oxon},
	isbn={1-134-15215-9}
}

@book{merton1968,
	author={Robert K. Merton},
	year={1968},
	title={Social theory and social structure},
	publisher={The Free Press},
	address={New York},
	isbn={0029211301}
}

@article{Chen2020,
	author={Lijia Chen and Pingping Chen and Zhijian Lin},
	year={2020},
	title={Artificial Intelligence in Education: A Review},
	journal={IEEE Access},
	volume={8},
	pages={75264-75278},
	isbn={2169-3536},
	doi={10.1109/ACCESS.2020.2988510}
}

@article{Ouyang2022,
author = {Ouyang, Fan and Zheng, Luyi and Jiao, Pengcheng},
title = {Artificial intelligence in online higher education: A systematic review of empirical research from 2011 to 2020},
year = {2022},
issue_date = {Jul 2022},
publisher = {Kluwer Academic Publishers},
address = {USA},
volume = {27},
number = {6},
issn = {1360-2357},
url = {https://doi.org/10.1007/s10639-022-10925-9},
journal = {Education and Information Technologies},
month = jul,
pages = {7893–7925},
numpages = {33}
}

@article{McDonald2025,
	author={Nora McDonald and Aditya Johri and Areej Ali and Aayushi Hingle Collier},
	year={2025},
	title={Generative artificial intelligence in higher education: Evidence from an analysis of institutional policies and guidelines},
	journal={Computers in Human Behavior: Artificial Humans},
	volume={3},
	pages={100121},
	isbn={2949-8821},
   url = {https://doi.org/10.1016/j.chbah.2025.100121
    }
}

@article{Chan2023,
	author={Cecilia Ka Yuk Chan},
	year={2023},
	title={A comprehensive {AI} policy education framework for university teaching and learning},
	journal={International Journal of Educational Technology in Higher Education},
	volume={20},
	number={1},
	pages={38-25},
	isbn={2365-9440},
	doi={10.1186/s41239-023-00408-3}
}

@article{Chatterjee2020,
  author  = {Chatterjee, Sheshadri and Bhattacharjee, Kalyan Kumar},
  title   = {Adoption of artificial intelligence in higher education: a quantitative analysis using structural equation modelling},
  journal = {Education and Information Technologies},
  volume  = {25},
  pages   = {3443--3463},
  year    = {2020},
  doi     = {10.1007/s10639-020-10159-7},
  url     = {https://doi.org/10.1007/s10639-020-10159-7}
}

@article{Tang2024,
	author={Xin Tang and Zhiqiang Yuan and Shaojun Qu},
	year={2025},
	title={Factors Influencing University Students' Behavioural Intention to Use Generative Artificial Intelligence for Educational Purposes Based on a Revised {UTAUT2} Model},
	journal={Journal of Computer Assisted Learning},
	volume={41},
	number={1},
	isbn={0266-4909},
	doi={10.1111/jcal.13105}
}

@article{Hussain2025,
  author  = {Hussain, Muhammad Munawar and Hanif, Shazia and Ghauri, Khadija and Ain, Qurat ul},
  title   = {The Role of Behavioral Intention in AI Adoption and Student Success in Higher Education Institutions: A UTAUT2 Perspective},
  journal = {Indus Journal of Social Sciences},
  volume  = {3},
  number  = {2},
  pages   = {341--357},
  year    = {2025},
  issn    = {2960-219X},
  url     = {https://doi.org/10.59075/ijss.v3i2.1221}
}

@article{Ma2024,
	author={Dongmin Ma and Huma Akram and I-H Chen},
	year={2024},
	title={Artificial Intelligence in Higher Education: A Cross-Cultural Examination of Students’ Behavioral Intentions and Attitudes},
	journal={International review of research in open and distance learning},
	volume={25},
	number={3},
	pages={134-157},
	isbn={1492-3831},
	doi={10.19173/irrodl.v25i3.7703}
}

@article{Zhou2024,
	author={Xue Zhou and Joanne Zhang and Ching Chan},
	year={2024},
	title={Unveiling students' experiences and perceptions of Artificial Intelligence usage in higher education},
	journal={Journal of university teaching \& learning practice},
	volume={21},
	number={6},
	pages={126-145},
	isbn={1449-9789},
	doi={10.53761/xzjprb23}
}

@article{Ursavas2025,
	author={Ömer Faruk Ursavaş and Yasin Yalçın and Hakan İslamoğlu and Eda Bakır-Yalçın and Mutlu Cukurova},
	year={2025},
	title={Rethinking the importance of social norms in generative {AI} adoption: investigating the acceptance and use of generative {AI} among higher education students},
	journal={International Journal of Educational Technology in Higher Education},
	volume={22},
	number={1},
	pages={38-22},
	isbn={2365-9440},
	doi={10.1186/s41239-025-00535-z}
}

@inproceedings{Yuan2025,
  author    = {Yuan, Qianshun},
  title     = {Game-Theoretic Analysis of the Collaborative Relationship Between Teachers and Students in the Application of ChatGPT in Higher Education},
  booktitle = {Proceedings of the 2025 IEEE International Conference on Education Innovation and Technology (ICEIT)},
  year      = {2025},
  pages     = {344--349},
  publisher = {IEEE},
  isbn      = {979-8-3315-4088-3},
  doi       = {10.1109/ICEIT64364.2025.10976210}
}

@article{You2023,
	author={Yanwei You and Yuquan Chen and Yujun You and Qi Zhang and Qiang Cao},
	year={2023},
	title={Evolutionary Game Analysis of Artificial Intelligence Such as the Generative Pre-Trained Transformer in Future Education},
	journal={Sustainability},
	volume={15},
	number={12},
	pages={9355},
	isbn={2071-1050},
	doi={10.3390/su15129355}
}

@article{AlZahrani2024Ethical,
  author  = {Al-Zahrani, Abdulrahman M. and Alasmari, Talal M.},
  title   = {Exploring the impact of artificial intelligence on higher education: The dynamics of ethical, social, and educational implications},
  journal = {Humanities and Social Sciences Communications},
  volume  = {11},
  pages   = {912},
  year    = {2024},
  url     = {https://doi.org/10.1057/s41599-024-03432-4}
}

@article{BerishaQehaja2025,
title = {Strategic integration of artificial intelligence solutions to transform teaching practices in higher education},
journal = {Foresight},
volume = {27},
number = {5},
pages = {970-991},
year = {2025},
issn = {1463-6689},
doi = {https://doi.org/10.1108/FS-04-2024-0079},
url = {https://www.sciencedirect.com/science/article/pii/S1463668925000161},
author = {Albana {Berisha Qehaja}},
}

@article{Ajjawi2020,
author = {Rola Ajjawi and David Boud},
title = {Examining the nature and effects of feedback dialogue},
journal = {Assessment \& Evaluation in Higher Education},
volume = {43},
number = {7},
pages = {1106--1119},
year = {2018},
publisher = {Routledge},
doi = {10.1080/02602938.2018.1434128},
URL = { https://doi.org/10.1080/02602938.2018.1434128
},
eprint = {     
        https://doi.org/10.1080/02602938.2018.1434128
    }
}

@article{MolloyBoudHenderson2020,
author = {Elizabeth Molloy and David Boud and Michael Henderson},
title = {Developing a learning-centred framework for feedback literacy},
journal = {Assessment \& Evaluation in Higher Education},
volume = {45},
number = {4},
pages = {527--540},
year = {2020},
publisher = {Routledge},
doi = {10.1080/02602938.2019.1667955},
URL = { 
    https://doi.org/10.1080/02602938.2019.1667955
    },
eprint = { 
            https://doi.org/10.1080/02602938.2019.1667955
    }
}

@article{bearman2022assessment,
	author={Margaret Bearman and Juuso Henrik Nieminen and Rola Ajjawi},
	year={2023},
	title={Designing assessment in a digital world: an organising framework},
	journal={Assessment and evaluation in higher education},
	volume={48},
	number={3},
	pages={291-304},
	isbn={0260-2938},
	doi={10.1080/02602938.2022.2069674}
}

@report{dawson2023assessmentAI,
  author       = {Lodge, Jason M. and Howard, Sarah and Bearman, Margaret and Dawson, Phillip and Associates},
  title        = {Assessment Reform for the Age of Artificial Intelligence},
  institution  = {Tertiary Education Quality and Standards Agency (TEQSA)},
  year         = {2023},
  month        = {September},
  address      = {Australia},
}

@Article{bittle2025genAIreview,
AUTHOR = {Bittle, Kyle and El-Gayar, Omar},
TITLE = {Generative AI and Academic Integrity in Higher Education: A Systematic Review and Research Agenda},
JOURNAL = {Information},
VOLUME = {16},
YEAR = {2025},
NUMBER = {4},
ARTICLE-NUMBER = {296},
URL = {https://www.mdpi.com/2078-2489/16/4/296},
ISSN = {2078-2489},
DOI = {10.3390/info16040296}
}

@article{kizilcec2024perceptions,
title = {Perceived impact of generative AI on assessments: Comparing educator and student perspectives in Australia, Cyprus, and the United States},
journal = {Computers and Education: Artificial Intelligence},
volume = {7},
pages = {100269},
year = {2024},
issn = {2666-920X},
doi = {https://doi.org/10.1016/j.caeai.2024.100269},
url = {https://www.sciencedirect.com/science/article/pii/S2666920X24000729},
author = {René F. Kizilcec and Elaine Huber and Elena C. Papanastasiou and Andrew Cram and Christos A. Makridis and Adele Smolansky and Sandris Zeivots and Corina Raduescu},
}

@Article{Khlaif2025Redesign,
AUTHOR = {Khlaif, Zuheir N. and Alkouk, Wejdan Awadallah and Salama, Nisreen and Abu Eideh, Belal},
TITLE = {Redesigning Assessments for AI-Enhanced Learning: A Framework for Educators in the Generative AI Era},
JOURNAL = {Education Sciences},
VOLUME = {15},
YEAR = {2025},
NUMBER = {2},
ARTICLE-NUMBER = {174},
URL = {https://www.mdpi.com/2227-7102/15/2/174},
ISSN = {2227-7102},
DOI = {10.3390/educsci15020174}
}

@article{Urquhart2026Assessment,
title = {Changing EAP assessment practices in the age of generative artificial intelligence: The case of Scottish higher education institutions},
journal = {Journal of English for Academic Purposes},
volume = {79},
pages = {101609},
year = {2026},
issn = {1475-1585},
doi = {https://doi.org/10.1016/j.jeap.2025.101609},
url = {https://www.sciencedirect.com/science/article/pii/S1475158525001407},
author = {Lewis Urquhart and Xuan Minh Ngo},
}

@article{Han2026CooperationWelfare,
title = {Cooperation versus social welfare},
journal = {Physics of Life Reviews},
volume = {56},
pages = {33-60},
year = {2026},
issn = {1571-0645},
doi = {https://doi.org/10.1016/j.plrev.2025.11.006},
url = {https://www.sciencedirect.com/science/article/pii/S1571064525001654},
author = {The Anh Han and Zhao Song and Theodor Cimpeanu and Manh Hong Duong and Marcus Krellner and Valerio Capraro and Matjaz Perc},
}

@inproceedings{NdidiZhaoTheAnh2025,
author = {Ogbo, Ndidi Bianca and Song, Zhao and Han, The Anh},
title = {Evolution of Coordination Through Institutional Incentives: An Evolutionary Game Theory Approach},
year = {2025},
isbn = {9798400722752},
publisher = {Association for Computing Machinery},
address = {New York, NY, USA},
url = {https://doi.org/10.1145/3772429.3772434},
doi = {10.1145/3772429.3772434},
booktitle = {Proceedings of the 2025 7th International Conference on Distributed Artificial Intelligence},
pages = {38–47},
numpages = {10},
location = {
},
series = {DAI '25}
}

\end{document}